\documentclass[pre,twocolumn,showpacs]{revtex4}

\usepackage{graphicx}
\usepackage{psfrag}
\usepackage{amsmath}   
\usepackage{amssymb}   
\usepackage{amsfonts}
\usepackage{bm}  
\usepackage{bbm}
\usepackage{mathrsfs}
\usepackage{mathbbol}
\usepackage{epsfig}
\usepackage{exscale}
\usepackage{lscape}
\usepackage{fancybox}
\usepackage{epic}

\usepackage{math4phy}

\usepackage{makeidx}   

\newcommand{\nn}{\nonumber}
\newcommand{\ve}{\vert}
\newcommand{\lk}{\left}
\newcommand{\rk}{\right}

\begin{document}

\title{Minimal length scales for the existence of local temperature}

\author{Michael Hartmann} 
\email{m.hartmann@imperial.ac.uk}
\affiliation{Institute of Mathematical Sciences, Imperial College London, 48 Princes' Gardens,
SW7 2BW, United Kingdom}

\affiliation{QOLS, Blackett Laboratory, Imperial College London, Prince Consort Road, London,
SW7 2BW, United Kingdom}

\date{\today}

\begin{abstract}
We review a recent approach to determine the minimal spatial length scales on which local
temperature exists. After mentioning an experiment where such considerations are of
relevance, we first discuss the precise definition of the existence of local temperature
and its physical relevance.
The approach to calculate the length scales in question considers homogenous chains of particles
with nearest neighbor interactions.
The entire chain is assumed to be in a thermal equilibrium state and it is analyzed when such
an equilibrium state at the same time exists for a local part of it.
The result yields estimates for real materials, the liability of which is discussed in the
sequel.
We finally consider a possibility to detect the existence or non-existence of a local
thermal state in experiment.
\end{abstract}

\maketitle


%
%
\section{Introduction}
\label{sec:intro}

Large systems in an equilibrium state may, despite their very large number of degrees of freedom,
be characterized by only very few quantities.
For example, an ideal gas is described by the simple ``thermal equation of state''
$p V = n k_B T$, where
$p$ is the pressure of the gas, $V$ its volume, $n$ the number of particles it contains, $T$
its temperature and $k_B$ Boltzmann's constant. In physics one refers to this kind of description
as a thermodynamical description.

How can such an extremely reduced description be justified? The reason why a
thermodynamical description works so well for equilibrium states is that,
with increasing number of particles in a
system, a dominant part of microstates have the same macroscopic properties.
Microstate refer here to a description where all degrees of freedom are specified.
As a result, thermodynamical behavior becomes ``typical''.

In a more mathematical language this fact is called the existence of the Thermodynamic Limit,
which merely means that intensive quantities such as the energy per particle approach
a limiting value that does no longer depend on the detailed configuration of the system as its
size increases. For example the energy per particle of a very large piece of solid does no longer
depend on whether this piece is lying on a table or is immersed in a bucket full of water,
provided it is in an equilibrium state, i.e. has the same temperature as its surrounding.

Obviously, in the Thermodynamic Limit, the difference between particles inside the solid and
those on the surface, which interact with the surrounding, becomes negligible.
That is why the size of the solid is important. Let us assume the solid had the shape of a sphere
and the density of the particles was uniform within it. Since the surface of a sphere with radius
$r$ is $4 \pi r^2$ and its volume is $(4 \pi / 3) r^3$, the ratio of particles sitting on the
surface over the total number
scales as $1 / r$ and thus becomes negligible as $r$ goes to infinity.
Such a type of scaling does apply not only to spheres but also to more general
geometries.

To analyze the existence of {\em local} temperatures, small parts of large systems are of
interest. These parts do inevitably interact with their surrounding.
For short range interactions between the constituent particles, again, only those particles
sitting on the boundary of the considered part interact with the environment.
Thus, the described scaling properties immediately give rise to the following
question%
: How large do parts of those systems have to be to permit a local
thermodynamical description, i.e. a thermodynamical description of the part alone?

For a long time, the problem,
besides being fundamental, may have been of purely academic interest, since thermodynamics
was only used to describe macroscopic systems, where deviations form the Thermodynamic Limit
may safely be neglected. However, with the advent of nanotechnology, the microscopic limit of
the applicability of thermodynamics became relevant for the interpretation of experiments
and may in the near future even have technological importance.

In recent years, amazing progress in the synthesis and processing of materials with structures on
nanometer length scales has been made \cite{Cahill2003,Williams1986,Varesi1998,Schwab2000}.
Experimental techniques have improved to such an extent that the measurement of thermodynamic
quantities like temperature with a spatial resolution on the nanometer scale seems within reach
\cite{Gao2002,Pothier1997,Aumentado2001}. 
To provide a basis for the interpretation of present day and future experiments in nanoscale
physics and technology and to obtain a better understanding of the limits of thermodynamics,
it is thus indispensable to clarify the applicability of thermodynamical concepts
on small length scales starting from the most fundamental theory at hand, i.~e. quantum mechanics.
In this context, one question appears to be particularly important and interesting:
Can temperature be meaningfully defined on nanometer length scales?

Why should we care about the non-existence of local temperature?
There are at least three situations for which this possibility needs special
attention: One obvious scenario refers to the limit of spatial resolution on 
which a temperature profile could be defined. However a spatially
varying temperature calls for non-equilibrium - a complication which we will exclude
here.
A second application concerns partitions on the nanoscale:
If a modular system in thermal equilibrium is partitioned into two pieces, say, the
two pieces need no longer be in a canonical state, let alone have the same local temperature.
Finally, local physical properties may show different behavior depending on whether the
local state
is thermal or not.

The existence of thermodynamical quantities, i.e. the existence of the
Thermodynamic Limit strongly
depends on the correlations between the considered parts of a system.
As mentioned above, with increasing diameter, the volume of a region in space grows
faster than its surface.
Thus effective interactions between two regions, provided they are short ranged,
become less relevant
as the sizes of the regions increase.
This scaling behavior is used to show that correlations between a region and its environment
become negligible in the limit of infinite region size and that therefore the Thermodynamic Limit exists
\cite{Fisher1964,Ruelle1969,Lebowitz1969}.

To explore the minimal region size needed for the application of thermodynamical concepts, situations far
away from the Thermodynamic Limit should be analyzed. On the other hand, effective correlations between
the considered parts need to be small enough \cite{Schmidt1998,Hartmann2003a}.

The scaling of interactions between parts of a system compared to the energy contained in the parts
themselves thus sets a minimal length scale on which correlations are
still small enough to permit the definition of local temperatures.
Here we review an approach to study this connection quantitatively
\cite{Hartmann2003b,Hartmann2004a}.


%
%
\section{Motivation: A Thermal Nanoscale Experiment}
\label{experiments}

In recent years, there has been substantial progress in the fabrication and operating
of material with structure on nanoscopic scales and nanoscale devices.
In this context, several experiments, that study thermal properties, have been done.
We describe here, as an example, one experiment that nicely shows where the
existence or non-existence of local temperature becomes relevant \cite{Kim2001}. 

%
%
%
\begin{figure}
\psfrag{C}{\hspace{-0.18cm} \raisebox{-0.05cm}{Nanotube}}
\psfrag{V}{\raisebox{-0.08cm}{\hspace{-0.2cm}$V\ind{h}$}}
\psfrag{Rh}{\hspace{-0.02cm}$R\ind{h}$}
\psfrag{Rs}{\hspace{-0.25cm} $R\ind{s}$}
\psfrag{Th}{\hspace{-0.25cm} $T\ind{h}$}
\psfrag{Ts}{$T\ind{s}$}
\includegraphics[width=7cm]{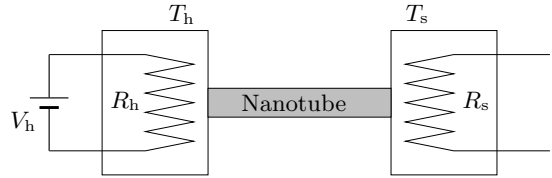}
\caption[Setup of the thermal nanoscale experiment.]{\label{nanowire_setup}
Setup of the experiment. Two, otherwise thermally well isolated islands
are connected by a carbon nanotube. The left island is heated by an electric current running
through the coil with resistance $R\ind{h}$ and thus maintained at the temperature $T\ind{h}$.
The temperature of the right island is measured via the temperature dependent
resistance $R\ind{s}$.}
\end{figure}

\begin{figure}
\psfrag{Rph}{${\bm R\ind{h}}$}
\psfrag{Rps}{\hspace{-0.25cm} ${\bm R\ind{s}}$}
\psfrag{N}{\hspace{-1.4cm} \raisebox{-0.0cm}{{\bf Carbon Nanotube}}}
\includegraphics[width=7cm]{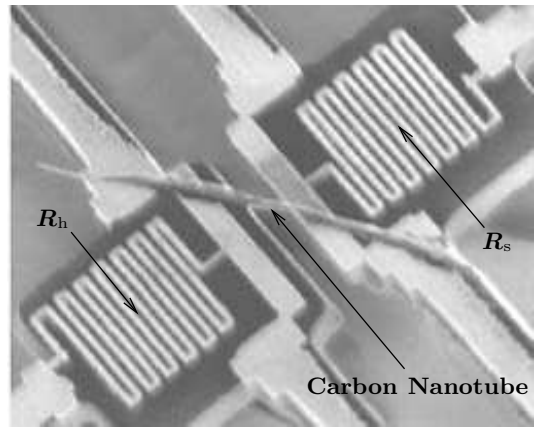}
\caption[Picture of the thermal nanoscale experiment.]{\label{nanowire_picture}
Picture of the setup. The heated island is in the lower left corner and
the island where the temperature is measured in the higher right corner.
Both are connected by a single carbon nanotube (With permission from A. Majumdar,
Nanoengineering Laboratory, University of California, Berkeley).}
\end{figure}

The experiment studies heat conduction across a carbon nanotube.
A sketch of the setup is given in figure \ref{nanowire_setup}.
Two, otherwise thermally well isolated islands are connected through a carbon nanotube of a
few $\mu$m length. \index{carbon nanotube}
One island is heated by an electric current that runs through a coil with the resistance
$R\ind{h}$. This island is thus at a ``hot'' temperature $T\ind{h}$. Heat can flow across
the nanotube to the other island, which is at a lower temperature $T\ind{s}$. This temperature
in turn is measured by another coil, the resistance of which $R\ind{s}$, depends on temperature.
Figure \ref{nanowire_picture} shows a picture of this setup.

%
%
%

\begin{figure}
\psfrag{C}{\hspace{-1.3cm} \raisebox{-0.0cm}{Carbon Nanotube}}
\psfrag{Th}{\hspace{-0.1cm} $T\ind{h}$}
\psfrag{Ts}{$T\ind{s}$}
\psfrag{T}{\hspace{-0.17cm} \raisebox{-0.0cm}{$T$}}
\psfrag{f}{\raisebox{-0.1cm}{{\Huge ?}}}
\includegraphics[width=7cm]{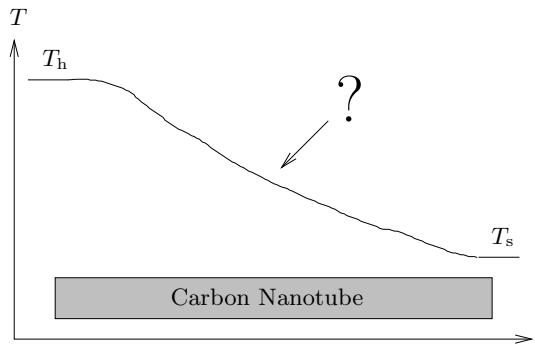}
\caption[Does a temperature profile exist or not?]{\label{nanowire_profile} The question,
whether a temperature profile exists for the
nanotube in the present setup, is not clarified.}
\end{figure}

At what point is the existence or non-existence of local temperatures of relevance for the
interpretation of this experiment?
We know that there is an electric current in the coil of the heated island.
This current constantly delivers thermal energy to the island. This energy is transported
across the nanotube to the other island, where we observe, that the temperature $T\ind{s}$ rises.
We thus know, that the nanotube connects a hot spot $T\ind{h}$ to a cold spot $T\ind{s}$.
This directly gives rise to the following questions: How hot is the nanotube in between?
Can we meaningfully talk at all about temperature of any part of the nanotube?
The answer to these two questions would clarify whether and in what sense a temperature profile
(see figure \ref{nanowire_profile}) could exist for the present setup.

%
%
%

While a temperature profile can obviously be defined and measured in a macroscopic version
of the present experiment, say two buckets of water at different temperatures and connected
via an iron bar, its existence, possible resolution and measurability are completely unclear
for the nanoscopic version.

To address this question,
we first discuss how to define the existence of local temperature in the
next section.


%
%
\section{What is Temperature?}
\label{temperature}

Temperature is one of the central quantities in thermodynamics and Statistical Mechanics.
Let us note here, that it is not a direct observable; it is not represented by an operator
in quantum mechanics (cf. sect. \ref{measure}). There exist two standard ways to define it:

\subsection{Definition in Thermodynamics}

The thermodynamical definition is purely empirical. Thermodynamics itself is an empirical
theory on systems whose macroscopic physics can be sufficiently characterized
by a set of a few variables
like volume, energy and the number of particles for example.
The values of this variables are called a macro state \cite{Adam1978}.
A system is said to be in equilibrium if its macro state is stationary for given
constraints.
As a consequence of this definition, a equilibrium state depends on the applied constraints,
e.g. whether the volume or the energy is kept constant etc.  
An important, special case of constraints, is a bipartite (or multipartite)
system with a fixed total energy,
where the parts may exchange energy among themselves.
The parts are then said to be in thermal equilibrium.
In thermodynamics, temperature is defined by the following property:

\bigskip

\begin{tabular}{rp{6cm}}
{\bf Definition:} &
{\em Two systems that can exchange energy and are in thermal equilibrium,
have the same temperature.}
\end{tabular}

\bigskip

To fix a temperature scale, a reference system is needed.
The simplest choice for this reference system is
the ideal gas (cf. sec. \ref{sec:intro}), where temperature may be defined by
\begin{equation}\label{thermo_temp_def}
T \equiv \frac{p \, V}{n \, k\ind{B}} \, .
\end{equation} 
Of course the above definition is unambiguous only if the states of thermal equilibrium form a
one-dimensional manifold \cite{Adam1978}.
Only then, a single parameter is sufficient for their characterization.
This parameter is the temperature $T$.

\subsection{Definition in Statistical Mechanics}

In statistical mechanics, temperature is defined via the derivative of the
entropy $S$ with respect to the
internal energy $\overline{E}$.
In quantum mechanics the entropy can be defined according to von Neumann
as
\begin{equation}\label{entropy_def}
S \equiv - k\ind{B} \Tr \op{\rho} \ln \op{\rho} \, ,
\end{equation} 
it is a measure of the amount of possible pure states, the system could be in.
With this definition, entropy always exists, but shows its standard properties, e.g. extensivity,
only in the Thermodynamic Limit \cite{Ruelle1969}.

In statistical mechanics, an equilibrium state is defined to be the state with the
maximal entropy, that is
the state with the maximal amount of accessible pure states.

For systems that interact with their surrounding,
such that they can exchange energy with it but have
a fixed expectation value for the energy, the equilibrium state is a so called canonical state,
described by a density matrix of the form
\begin{equation} \label{can_def}
\op{\rho} = \frac{\exp ( - \beta \op{H})}{Z} \, ,
\end{equation} 
where the partition sum\index{partition sum} $Z$ normalizes $\op{\rho}$ such that
$\Tr \op{\rho} = 1$.

In quantum mechanics the internal energy is given by the expectation value of the
energy,
\begin{equation}\label{int_en_def}
\overline{E} \equiv \Tr \op{\rho} \op{H} \, ,
\end{equation} 
where the Hamiltonian is the energy operator of the isolated system at hand.
It does not contain any interactions of the system with its environment.
The internal energy is therefore
a property of the system itself, it only depends on the state of the system and
not on the state of the
environment.

Temperature is then defined by
\begin{equation}\label{temp_def}
\frac{1}{T} \equiv \pop{S}{\overline{E}} \, ,
\end{equation} 
which in turn exists as long as the entropy $S$ is a function of the internal energy
$\overline{E}$.
However, the notion of temperature, as defined in equation (\ref{temp_def}),
just like entropy shows its
characteristic thermodynamical properties (see above) only for equilibrium states
\cite{Adam1978,Ruelle1969,Tolman1967}.

\subsection{Local Temperature}

Local temperature is, by definition, the temperature of a part of a larger system.
Hence, this subsystem is not isolated but can exchange energy with its surrounding.
On the other hand we limit our considerations to cases without particle exchange.
Hence, the following convention appears to be reasonable:

\bigskip

\begin{tabular}{rp{6cm}}
{\bf Definition:} &
{\em Local temperature exists if the considered part of the system is in a canonical state.}
\end{tabular}

\bigskip

Note: While a local state can always uniquely be defined by tracing out the rest of the system,
this definition calls, in addition, for the (approximate) existence of some local spectrum.

Besides being based on statistical mechanics
there are further practical reasons for this definition:
The canonical distribution is an exponentially decaying function of energy characterized by one
single parameter, temperature.
This implies that there is a one to one mapping between temperature and
the expectation values of observables, by which it is usually measured.
Temperature measurements
via different observables thus yield the same result, contrary to distributions with several
parameters. 

This is a basic property of systems that can be characterized by
thermodynamic description.
The temperature, if it exists, describes a system in a sufficiently complete way,
such that several
properties of it can be predicted if one only knows its temperature (see also sec. \ref{measure}).

\begin{figure}[h]
\psfrag{eta}{\small \hspace{-0.35cm} \raisebox{0.5cm}{$\eta (E)$}}
\psfrag{pb}{\small \hspace{0.72cm} \raisebox{0.05cm}{$\bra{\varphi} \rho \ket{\varphi}$}}
\psfrag{prod}{\small \hspace{0.12cm} \raisebox{0.1cm}{$\eta (E) \cdot \bra{\varphi}
\rho \ket{\varphi}$}}
\psfrag{E1}{\small \hspace{-0.4cm} $E\ind{min}$}
\psfrag{E2}{\small \hspace{-0.23cm} $E\ind{max}$}
\psfrag{E}{\small \hspace{-0.5cm} \raisebox{-0.75cm}{$\overline{E}$}}
\psfrag{n}{}
\psfrag{c1}{$\: E$}
\includegraphics[width=7cm]{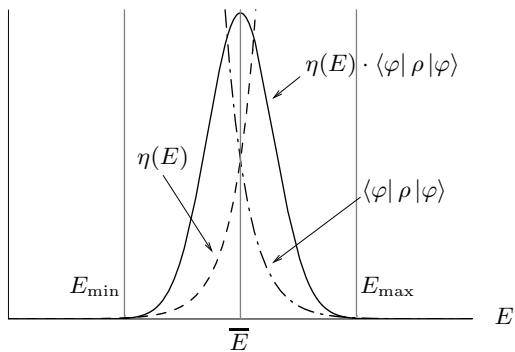}
\caption[Density of states, canonical distribution and their product for a
typical many body system.]{The
product of the density of states $\eta (E)$ times the occupation probabilities
$\bra{\varphi} \rho \ket{\varphi}$ forms a strongly pronounced peak at $E = \overline{E}$.}
\label{peak}
\end{figure}

Why does the distribution need to be exponentially decaying?
In large systems with a modular structure, the density of states is a strongly growing
function of energy \cite{Tolman1967}. The product of the density of states times an
exponentially decaying distribution of occupation probabilities will thus form a strongly
pronounced peak at the internal energy $\overline{E}$, see figure \ref{peak}.

%
%
%

If the distribution was not exponentially decaying,
the product of the density of states times the distribution would not have a pronounced peak
and thus physical quantities like energy could not have ``sharp'' values.


%
%
%
\section{General Theory for the Existence of Local Temperature}
\label{general}

After having introduced and discussed the conception, when temperature is defined to exist
locally, we now turn to describe the approach to analyze its existence.

Since temperature is defined to exist locally, i.e. for a given part of
the system we consider, if the respective part is in a thermal equilibrium state,
it is defined to exist on a certain length scale, if all possible
partitions of the corresponding size are simultaneously in an equilibrium state.

This requirement for the local equilibrium states in the parts
to exist at the same time needs some
further discussion: For a given temperature profile, it should not make a difference
whether the profile is scanned by one single thermometer, which is moved in small steps
across the sample, or whether the profile is measured by several thermometers
simultaneously, which are located at small distances to each other.

For systems which are globally in a non-equilibrium state it is very difficult to
decide under what conditions equilibrium states show up locally \cite{Kreuzer1984} and
only very few exact results are known \cite{Meixner1941}.
Nonetheless, whenever local equilibrium exists, the macroscopic temperature gradient is small
($\delta T / T \ll 1$).
Here, we restrict ourselves to systems which are in a global equilibrium state (\ref{can_def}).
In these situations, subunits
of the total system are in an equilibrium state whenever their effective interaction
is weak enough and correlations between them are small so that
the global thermal state approximately factorizes into a product of local thermal states.

Whenever the macroscopic temperature gradient is small ($\delta T / T \ll 1$),
one would expect the results to be applicable even for situations with only local equilibrium
but non-equilibrium on the global scale.

To explore how local temperature can exist, that is how small the respective part may be,
one needs look at parts of
different sizes. The idea behind this approach is the scaling behavior which ensures
the existence of the Thermodynamic Limit (cf. section \ref{sec:intro})
\cite{Ruelle1969,Lebowitz1969}:

We consider systems that are composed of elementary subsystems with short range interaction, for
simplicity say nearest neighbor interaction.
If then $n$ adjoining subsystems form a part, the energy of the part is $n$
times the average energy per
subsystem and is thus expected to grow as the size of the part, $n$.
Since the subsystems only interact with their nearest neighbors, two adjacent parts
interact via the two subsystems at the respective boundaries, only.
As a consequence, the effective coupling between two parts
is independent of the part size $n$ and thus becomes less relevant
compared to the energy contained in the
parts as their size increases.

\subsection{The Model}

As models we consider here homogeneous (i.e. translation invariant) systems with nearest neighbor
interactions which we divide into identical parts. The Hamiltonian of the system thus reads
\begin{equation}\label{hamil}
H = \sum_{i} H_i + I_{i,i+1} \, ,
\end{equation}
where the index $i$ labels the elementary subsystems. $H_i$ is the Hamiltonian of subsystem $i$,
$I_{i,i+1}$ the interaction between subsystem $i$ and $i+1$ and
periodic boundary conditions are assumed.

Now $N_G$ groups of $n$ subsystems each
(index $i \rightarrow (\mu-1) n + j; \: \mu = 1, \dots, N_G; \: j = 1, \dots, n$)
are formed and the Hamiltonian is split into two parts,
\begin{equation}
\label{hsep}
H = H_0 + I \, ,
\end{equation}
where $H_0$ is the sum of the Hamiltonians of the isolated groups,
\begin{eqnarray}\label{isogroups}
H_0 & = & \sum_{\mu=1}^{N_G} \mathcal{H}_{\mu} \qquad
\text{with} \\
\mathcal{H}_{\mu} & = & \sum_{j=1}^n H_{n (\mu - 1) + j} + \sum_{j=1}^{n-1} I_{n (\mu-1) + j,\, n (\mu-1) + j + 1} \, , \nn
\end{eqnarray}
and $I$ contains the interaction terms of each group with its neighbor group,
\begin{equation}\label{eq:3}
I = \sum_{\mu=1}^{N_G} I_{\mu n,\mu n + 1} \, .
\end{equation}
\begin{figure}
\vspace{0.2cm}
\psfrag{n}{\hspace{-0.1cm}\raisebox{0.01cm}{$n$}}
\includegraphics[width=7cm]{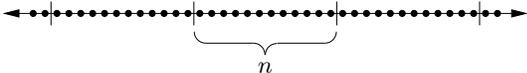}
\caption{Groups of $n$ adjoining subsystems are formed.}
\label{chain}
\end{figure}
The eigenstates $\ket{a}$ of the Hamiltonian $H_0$,\linebreak $H_0 \, \ket{a} = E_a \, \ket{a}$,
are products of group eigenstates of the individual groups,
\begin{equation}
\label{prodstate2}
\ket{a} = \prod_{\mu = 1}^{N_G} \otimes \ket{a_{\mu}} \qquad \text{with} \qquad
\mathcal{H}_{\mu} \ket{a_{\mu}} = E_{\mu} \ket{a_{\mu}} \, ,
\end{equation}
where $E_{\mu}$ is the energy of one subgroup only and\linebreak
$E_a = \sum_{\mu=1}^{N_G} E_{\mu}$.

\subsection{Thermal State in the Product Basis}

To test whether a part $H_{\mu_0}$ is in a thermal state,
we have to calculate its reduced
density matrix by tracing out the rest of the system. This trace can only be performed in
the basis formed by the states $\ket{a}$,~(\ref{prodstate2}).
We thus have to write the global equilibrium state (\ref{can_def}) in this basis.
Denoting the eigenstates and eigenenergies of the global Hamiltonian
with Greek indices, $\ket{\varphi}$,  $\ket{\psi}$ and $E_{\varphi}$, $E_{\psi}$,
the global equilibrium state $\op{\rho}$ reads
\begin{equation}
\bra{\varphi} \op{\rho} \ket{\psi} = \frac{\ee^{- \beta E_{\varphi}}}{Z} \: \delta_{\varphi \psi}
\end{equation}
in the global eigenbasis and the diagonal elements in the product basis are 
\begin{equation}
\label{eq:1}
\bra{a} \op{\rho} \ket{a} =
\int_{E_0}^{E_1} w_a (E) \: \frac{\ee^{- \beta E}}{Z} \: dE \: ,
\end{equation}
where the original sum $\sum_{\varphi}$ has been replaced by an integral over the energy.

$w_a (E)$ is the probability to obtain an energy value between $E$ and $E + \Delta E$ if the
total energy $H$ is measured for as system in the state $\ket{a}$, i.e.
\begin{equation}
w_a (E) = \frac{1}{\Delta E} \,
\sum_{\{ \ket{\varphi}: E \le E_{\varphi} < E + \Delta E \}}
\lk| \sprod{a}{\varphi} \rk|^2 \, .
\end{equation}
where the sum runs over all states $\ket{\varphi}$ with energy eigenvalues
$E_{\varphi}$ in the respective energy range and $\Delta E$ is small.
$E_0$ is the energy of the ground state and $E_1$ the upper limit of the spectrum,
which should be taken to be infinite if the spectrum does not have an upper bound.

We thus have to know the distributions $w_a (E)$ in order to be able to compute the
reduced density matrices of the groups and to test whether they are of canonical form.

Fortunately, one can indeed show that there exists a quantum central limit theorem
for many particle
systems with nearest neighbor interactions \cite{Hartmann2003,Hartmann2004b}.
Therefore, in the limit of infinitely many groups,
$w_a$ takes on the following form:
\begin{equation} \label{prob_dens}
\lim_{N_G \to \infty} w_a (E) =
\frac{1}{\sqrt{2 \pi} \, \Delta_a} \: \exp \lk( - \frac{\lk(E - \overline{E}_a \rk)^2}
{2 \, \Delta_a^2} \rk) \, ,
\end{equation}
where $\overline{E}_a$, the expectation value of $H$ in the state $\ket{a}$, and
$\Delta_a^2$, its variance, read
\begin{eqnarray}
\overline{E}_a & \equiv & \bra{a} H \ket{a} \qquad \text{and} \label{expect_v}\\
\Delta_a^2 & \equiv & \bra{a} H^2 \ket{a} - \bra{a} H \ket{a}^2 \, .\label{squared_ w}
\end{eqnarray}
Note here, that the limit of infinite number of groups is taken while the size of each
individual group remains finite.

For the theorem to hold, two further conditions have to be met: The energy of each group
including its interactions with the neighboring group has to be bounded and the variance
$\Delta_a^2$ has to grow faster than $N_G C$ for some positive constant $C$.
In scenarios, where the energy spectrum of each elementary subsystem has an upper limit,
such as spins, the first condition is met a priori.
For subsystems with an infinite energy spectrum, such as harmonic oscillators,
the present analysis is restricted to states where the energy of every group,
including the interactions with its neighbor groups, is bounded.
Thus, the considerations do not apply
to product states $\ket{a}$, for which all the energy was located in only one
group or only a small
number of groups. The number of such states is vanishingly small compared
to the number of all product states.

The expectation value of the entire Hamiltonian $H$ in the state $\ket{a}$, $\overline{E}_a$,
is the sum of the
energy eigenvalue of the isolated groups $E_a$ and a term that contains the interactions,
\begin{equation}
\overline{E}_a = E_a + \varepsilon_a \, ,
\end{equation}
Therefore, the two quantities $\varepsilon_a$ and $\Delta_a^2$ can also be
expressed in terms of the
interaction (see eq. (\ref{hsep})) only,
\begin{eqnarray}
\varepsilon_a & = & \bra{a} I \ket{a} \qquad \text{and} \label{eq:2}\\
\Delta_a^2 & = & \bra{a} I^2 \ket{a} - \bra{a} I \ket{a}^2 \label{eq:4}\, ,
\end{eqnarray}
meaning that $\varepsilon_a$ is the expectation value and $\Delta_a^2$ the squared
width of the interactions
\index{squared width}
in the state $\ket{a}$. Note that $\varepsilon_a$ has a classical counterpart while
$\Delta_a^2$ is purely
quantum mechanical.
It appears because the commutator $[H,H_0]$ is nonzero, and the distribution $w_a(E)$
therefore has nonzero
width. 

Applying equation (\ref{prob_dens}) to calculate the integral in equation (\ref{eq:1})
yields for $N_G  \gg 1$,
\begin{equation}
\label{newrho2}
\begin{split}
\bra{a} \op{\rho} \ket{a} = \frac{1}{Z} \,
\exp \lk(- \beta \lk(E_{a} + \varepsilon_a\rk) + \frac{\beta^2 \Delta_a^2}{2} \rk)
\times \\[0.3cm]
\times \, \frac{1}{2} \,
\lk[\text{erfc} \lk( \frac{E_0 - E_{a} - \varepsilon_a + \beta \Delta_a^2}{\sqrt{2} \, \Delta_a} \rk) - \rk. \\[0.3cm]
\lk. 
\text{erfc} \lk( \frac{E_1 - E_{a} - \varepsilon_a + \beta \Delta_a^2}{\sqrt{2} \, \Delta_a} \rk) \rk] \, ,
\end{split}
\end{equation}
where $\text{erfc} (x)$ is the conjugate Gaussian error function \cite{Abramowitz1970},
%
%
The second error function appears only if the energy is bounded and the
integration extends from the energy of the ground state $E_0$ to the upper
limit of the spectrum $E_1$.

Note that the arguments of the conjugate error functions grow proportional to $\sqrt{N_G}$
or stronger, therefore the asymptotic expansion of the latter \cite{Abramowitz1970}
may be used for $N_G \gg 1$.

The off diagonal elements $\bra{a} \hat \rho \ket{b}$ vanish for
$\lk| E_a - E_b \rk| > \Delta_a + \Delta_b$ because the overlap of the two Gaussian
distributions becomes
negligible. For $\lk| E_a - E_b \rk| < \Delta_a + \Delta_b$, the transformation
involves an integral over frequencies and thus these terms are significantly smaller than
the entries on the diagonal.

\subsection{Conditions for Local Thermal States}

We now test under what conditions the density matrix $\hat \rho$ may be approximated by a product
of canonical density matrices with temperature $\beta\ind{loc}$ for each subgroup $\mu = 1, 2, \dots, N_G$.
\index{canonical state}
Since the trace of a matrix is invariant under basis transformations, it is sufficient to verify
the correct energy dependence of the product density matrix.
If we assume periodic boundary conditions,\index{product density matrix}
all reduced density matrices are equal and if they were canonical their product
would be of the form\linebreak $\bra{a} \op{\rho} \ket{a} \propto \exp(- \beta\ind{loc} E_a)$.
We thus have to verify whether the logarithm of rhs of equation (\ref{newrho2})
is a linear function of the energy $E_a$ defined in equation (\ref{prodstate2}) and below,
\begin{equation} \label{log}
\ln \lk( \bra{a} \op{\rho} \ket{a} \rk) \approx - \beta\ind{loc} \, E_a + c \, ,
\end{equation}
where $\beta\ind{loc}$ and $c$ are constants.

Applying the asymptotic expansion of the conjugate error function to (\ref{newrho2}) shows that
equation (\ref{log}) can only be true for
\begin{eqnarray}
\frac{E_a + \varepsilon_a  - E_0}{\sqrt{N_G} \, \Delta_a} & > & \beta \frac{\Delta_a^2}{\sqrt{N_G} \, \Delta_a} \qquad \text{and} \label{cond_const}\\[0.2cm]
- \varepsilon_a + \frac{\beta}{2} \, \Delta_a^2 & \approx & c_1 E_a + c_2 \, , \label{cond_linear_1}
\end{eqnarray}
where $c_1$ and $c_2$ are constants.
These two conditions constitute the general result of this section.

Note that $\varepsilon_a$ and $\Delta_a^2$ need not be functions of $E_a$ and therefore in general
cannot be expanded in a Taylor series.

Temperature becomes intensive, if the constant $c_1$ vanishes, \index{intensivity}
\begin{equation} \label{intensivity}
\lk| c_1 \rk| \ll 1 \qquad \Rightarrow \qquad \beta\ind{loc} = \beta \, .
\end{equation}
Even if this was not the case, temperature might still exist locally.

For the existence of local temperature, one should only require that the
diagonal elements (\ref{eq:1}) are
canonically distributed in an appropriate energy range, $E\ind{min} \le E_a \le E\ind{max}$.
As described in the previous section, the density of states $\eta (E)$ is, for
large modular systems, an exponentially growing function of energy and its product with the
exponentially decaying canonical distribution $\bra{\varphi} \rho \ket{\varphi}$ forms a
strongly pronounced peak at the expectation value of the global energy $\overline{E}$,
see figure \ref{peak}.

If the diagonal elements (\ref{eq:1}) are canonically distributed in an energy range,
that is centered at this peak and is large enough to entirely cover it, all observables
with non-vanishing matrix elements in that range show the same behavior as for a
true canonical distribution.
Observables which are not of that kind are in general not of interest.
If one considers for example 1 kg of iron at 300 Kelvin with an average energy of
roughly 130 kJ, one is usually not interested in processes,
that take place at energies of 0.1 kJ or $10^5$ kJ.

Therefore a pertinent and ``safe'' choice for the energy range
$E\ind{min} \le E_a \le E\ind{max}$ should be
\begin{equation} \label{e_range}
\begin{array}{rcl}
{\displaystyle E\ind{min}} & = & {\displaystyle \text{max}
\lk( \lk[E_a\rk]\ind{min} \, , \,
\frac{1}{\alpha} \overline{E} + E_0 \rk)}\\[0.5cm]
{\displaystyle E\ind{max}} & = & {\displaystyle \text{min}
\lk( \lk[E_a\rk]\ind{max} \, , \, \alpha
\overline{E} + E_0 \rk) \, ,}
\end{array}
\end{equation}
where $\alpha \gg 1$ and $\overline{E}$ will in general depend on the global temperature $\beta$.
In equation (\ref{e_range}), $\lk[E_{\mu}\rk]\ind{min}$ and $\lk[E_{\mu}\rk]\ind{max}$ denote
the minimal and maximal values $E_{\mu}$ can take on.

%
%
%
\begin{figure}[h]
\bigskip

\psfrag{3}{\small \hspace{+0.2cm} $ $}
\psfrag{4}{\small \hspace{+0.2cm} $ $}
\psfrag{5}{\small \raisebox{-0.1cm}{
$ $}}
\psfrag{6}{\small \raisebox{-0.1cm}{
$ $}}
\psfrag{7}{\small \raisebox{-0.1cm}{
$ $}}
\psfrag{8}{\small \raisebox{-0.1cm}{
$ $}}
\psfrag{E1}{\small \hspace{-0.32cm} $E\ind{low}$}
\psfrag{E2}{\small \hspace{-0.24cm} $E\ind{high}$}
\psfrag{n}{\hspace{-0.7cm} \raisebox{0.1cm}{$\ln \lk( \bra{a} \rho \ket{a} \rk)$}}
\psfrag{c1}{$\: E$}
\includegraphics[width=7cm]{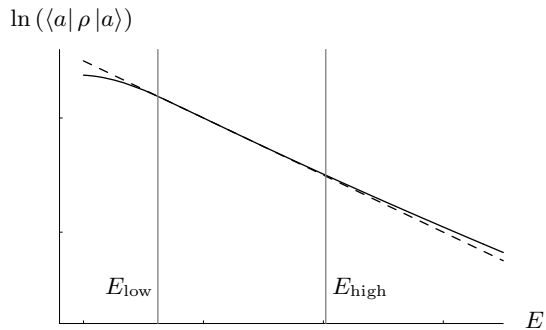}
\caption{
$\ln ( \langle a | \rho | a \rangle )$
for $\rho$ as in equation (\ref{newrho2}) (solid line) and a canonical density matrix $\rho$ (dashed line) for a harmonic chain.}
\label{visual}
\end{figure}

Figure \ref{visual} shows the logarithm of equation (\ref{newrho2}) and the logarithm of a
canonical distribution with the same $\beta$ for the example of a harmonic chain. The actual
density matrix is more mixed than the canonical one.
In the interval between the two vertical lines, both criteria
(\ref{cond_const}) and (\ref{cond_linear_1}) are satisfied.
For $E < E\ind{low}$ (\ref{cond_const}) is violated and (\ref{cond_linear_1}) for\linebreak
$E > E\ind{high}$. To allow
for a description by means of canonical density matrices, the
group size needs to be chosen such that $E\ind{low} < E\ind{min}$ and $E\ind{high} > E\ind{max}$.

For a model of the class considered here, the two conditions
(\ref{cond_const}) and (\ref{cond_linear_1})
must both be satisfied. In the following sections, these fundamental criteria will be applied
to a concrete model.


%
%
\section{Harmonic Chain} 
\label{harmonicchain}

We consider a harmonic chain of $N_G \cdot n$ particles of mass $m$ and
spring constant $\sqrt{m} \, \omega_0$ \cite{Hartmann2003b,HartmannHandb}.
In this case, the respective terms in the Hamiltonian (\ref{hamil}) read
\begin{eqnarray}
H_i & = & \frac{m}{2} \, p_i^2 + \frac{m}{2} \, \omega_0^2 \, q_{i}^2 \\
I_{i, i+1} & = & - m \, \omega_0^2 \, q_{i} \, q_{i+1} \, ,
\end{eqnarray}
where $p_i$ is the momentum of the particle at site $i$ and $q_{i}$ the displacement
from its equilibrium position $i \cdot a_0$ with $a_0$ being the distance
between neighboring particles at equilibrium.
We divide the chain into $N_G$ groups of $n$ particles each and thus get
a partition of the type considered in section \ref{general}.

The Hamiltonian of each group is diagonalized by a Fourier transform and
the definition of creation
and annihilation operators $a_{k}^{\dagger}$ and $a_{k}$ for the Fourier modes
\cite{Hartmann2004a}. In this way we get
\begin{equation}
E_a = \sum_{\mu=1}^{N_G} E_{\mu} \quad
\text{with} \quad E_{\mu} = \sum_{k} \omega_k \left( n_k^a (\mu) + \frac{1}{2} \right) \, ,
\end{equation}
where $k = \pi l / (a_0 \, (n+1))$ $(l = 1, 2, \dots, n)$ and the
frequencies $\omega_k$ are given by
$\omega^2_{k} = 4 \, \omega_0^2 \, \sin^2\lk(k \, a_0 / 2\rk)$.
$n_k^a (\mu)$ is the occupation number of mode $k$ of
group $\mu$ in the state $\ket{a}$. We choose units, where $\hbar = 1$.
Let us first verify, that the central limit theorem (\ref{prob_dens}) applies to this model,
i.e. that the required assumptions (given below equation (\ref{squared_ w})) are met.

To see that $\Delta_a^2$ grows faster than $N_G C$ $(C > 0)$, one needs to
express the group interaction $V(q_{\mu n}, q_{\mu n + 1})$ in terms of
$a_{k}^{\dagger}$ and $a_{k}$; this yields $\Delta_a^2 = \sum_{\mu=1}^{N_G} \Delta_{\mu}^2$
where $\Delta_{\mu}^2 > 0$, implying that the assumption is met.

Since the spectrum of every single oscillator is infinite, the requirement that the energy per
group should be bounded can only
be satisfied for states, for which the energy of the system
is distributed among a relevant fraction
of the groups. As discussed in section \ref{general}, states where this is not the case
constitute only a negligible fraction of all product states $\ket{a}$.

The expectation values of the group interactions (eq.(\ref{eq:2})) vanish, $\varepsilon_a = 0$,
while the widths
$\Delta^2_{\mu}$ according to equation (\ref{squared_ w})
depend on the occupation numbers $n_{k}(\mu)$ and therefore on the energies
$E_{\mu}$.
To analyze conditions (\ref{cond_const}) and (\ref{cond_linear_1}),
one makes use of the continuum or Debye approximation \cite{Kittel1983}, requiring
$n \gg 1$, $a_0 \ll l$, where $l = n \, a_0$, and the length of the chain to be finite.
As will become clear below,
the resulting minimal group sizes $n\ind{min}$ are larger than $10^3$ for all temperatures and
the application of the Debye approximation is well justified.

Using this approximation we now have $\omega_k = v \, k$ with the constant velocity of sound
$v = \omega_0 \, a_0$ and the width of the group interaction reads
\begin{equation}
\label{harmsigma_2}
\Delta^2_{\mu} = \frac{4}{n^2} \, E_{\mu} \, E_{\mu+1} \, ,
\end{equation}
where $n+1 \approx n$ has been used.

The relevant energy scale is introduced by the thermal expectation
value of the entire chain
\begin{equation}
\label{intenergy}
\overline{E} = E_0 \, + \, N_G n k_B \Theta \left( \frac{T}{\Theta} \right)^2
\int_0^{\Theta / T} \frac{x}{\ee^x - 1} \, dx \, ,
\end{equation}
and the ground state energy $E_0$ is given by
\begin{equation}
\label{groundenergy}
E_0 = N_G n k_B \Theta \left( \frac{T}{\Theta} \right)^2
\int_0^{\Theta / T} \frac{x}{2} \, dx = \frac{N_G n k_B \Theta}{4} \, ,
\end{equation}
where $\Theta$ is the Debye temperature \cite{Kittel1983}\index{Debye temperature}.

Inserting equation (\ref{intenergy}) and (\ref{groundenergy}) into equations
(\ref{cond_const}) and (\ref{cond_linear_1}), taking into account (\ref{e_range}),
one can now calculate the minimal $n$ for given $\alpha,\Theta$ and $T$.
In doing so, one needs to introduce another accuracy parameter $\delta$ which, for the
rhs of eq. (\ref{cond_linear_1}), quantifies how much smaller terms quadratic and
higher order in $E_a$ are compared to the zero order and linear ones.
More precisely, $\delta$ is the ratio of the higher order terms to the (at most) linear ones.
 
\begin{figure}[h]
\psfrag{-4.1}{\small \raisebox{-0.12cm}{$10^{-4}$}}
\psfrag{-3.1}{\small \raisebox{-0.12cm}{$10^{-3}$}}
\psfrag{-2.1}{\small \raisebox{-0.12cm}{$10^{-2}$}}
\psfrag{-1.1}{\small \raisebox{-0.12cm}{$10^{-1}$}}
\psfrag{1.1}{\small \raisebox{-0.12cm}{$10^{1}$}}
\psfrag{2.1}{\small \raisebox{-0.12cm}{$10^{2}$}}
\psfrag{3.1}{\small \raisebox{-0.12cm}{$10^{3}$}}
\psfrag{4.1}{\small \raisebox{-0.12cm}{$10^{4}$}}
\psfrag{1}{}
\psfrag{2}{\small \hspace{+0.2cm} $10^{2}$}
\psfrag{3}{}
\psfrag{4}{\small \hspace{+0.2cm} $10^{4}$}
\psfrag{5}{}
\psfrag{6}{\small \hspace{+0.2cm} $10^{6}$}
\psfrag{7}{}
\psfrag{8}{\small \hspace{+0.2cm} $10^{8}$}
\psfrag{n}{\raisebox{0.1cm}{$n\ind{min}$}}
\psfrag{c1}{$\: T / \Theta$}
\includegraphics[width=7cm]{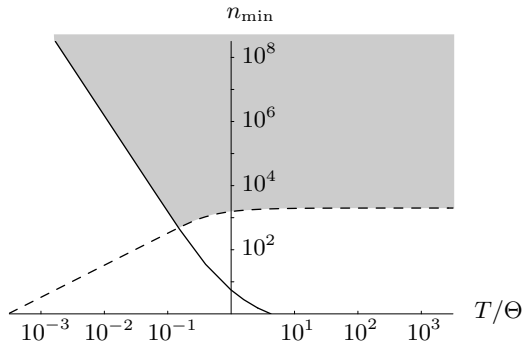}
\caption[$n\ind{min}(T)$ for a harmonic chain.]{$n\ind{min}$ as a function of $T / \Theta$
for a harmonic chain as determined by eq. (\ref{cond_const})
(solid line) and as determined by eq. (\ref{cond_linear_1}) (dashed line)
for $\alpha = 10$ and $\delta = 0.01$. Local temperature exists in the shaded region.}
\label{temp}
\end{figure}

Figure \ref{temp} shows $n\ind{min}$ for $\alpha = 10$ and $\delta = 0.01$
as a function of $T / \Theta$.

%
%
%
%

For high (low) temperatures $n\ind{min}$ can be estimated by
\begin{equation} \label{asympnmin}
n\ind{min} \approx \left\{
\begin{array}{lcr}
{\displaystyle \frac{2 \, \alpha}{\delta}} & {\displaystyle \text{for}} &
{\displaystyle T > \Theta}\\[0.3cm]
{\displaystyle \frac{3 \alpha}{2 \, \pi^2} \: \frac{\Theta^3}{T^3}} &
{\displaystyle \text{for}} & {\displaystyle T < \Theta}
\end{array}
\right.
\end{equation}
In addition, local temperatures are equal to the global one
whenever they exist, $\beta\ind{loc} = \beta$, implying that temperature is intensive
(see eq. (\ref{intensivity})).

In the following section the results obtained above will be applied to real
materials.


%
%
\section{Estimates for Real Materials}
\label{real}

Thermal properties of insulating solids can successfully be described by harmonic lattice models.
Probably the best known example of such a successful modeling is the correct prediction
of the temperature dependence of the specific heat based on the Debye theory \cite{Kittel1983}.
Therefore one would expect the present  approach to give reasonable estimates for real
materials, too. 

We thus take the results obtained in section \ref{harmonicchain} for the harmonic chain
and insert the corresponding parameters, in particular the Debye temperature
which can be found tabulated \cite{Kittel1983}. One obtains a length scale by multiplying
$n\ind{min}$ with the corresponding lattice constant.
The minimal length scale on which intensive temperatures exist in insulating solids should
thus given by 
\begin{equation}
\label{length}
l\ind{min} = n\ind{min} \, a_0,
\end{equation}
where $a_0$ is the lattice constant, the distance between neighboring atoms.

\begin{figure}[h]
\psfrag{-1.1}{\small \raisebox{-0.1cm}{$10^{-1}$}}
\psfrag{1.1}{\small \raisebox{-0.1cm}{$10^{1}$}}
\psfrag{2.1}{\small \raisebox{-0.1cm}{$10^{2}$}}
\psfrag{3.1}{\small \raisebox{-0.1cm}{$10^{3}$}}
\psfrag{4.1}{\small \raisebox{-0.1cm}{$10^{4}$}}
\psfrag{5.1}{\small \raisebox{-0.1cm}{$10^{5}$}}
\psfrag{2}{\small \hspace{-0.7cm} $10^{-8}$}
\psfrag{4}{\small \hspace{-0.7cm} $10^{-6}$}
\psfrag{6}{\small \hspace{-0.7cm} $10^{-4}$}
\psfrag{8}{\small \hspace{-0.7cm} $10^{-2}$}
\psfrag{10}{\small \hspace{-0.0cm} $1$}
\psfrag{n}{\raisebox{0.1cm}{\hspace{-0.2cm}$l\ind{min}$ [m]}}
\psfrag{c1}{$\quad T$ [K]}
\hspace{-0.3cm}
\includegraphics[width=7cm]{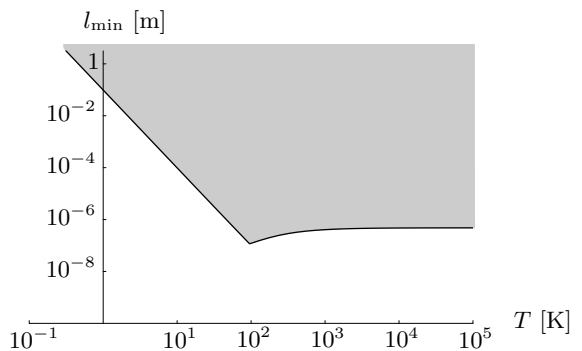}
\caption[$l\ind{min} (T)$ for crystalline silicon.]{$l\ind{min}$ as a function of temperature $T$
for crystalline silicon. $a_0 \approx 2.4 \,${\AA},
$\Theta \approx 645 \,$K, accuracy parameters $\alpha=10$ and $\delta=0.01$.
Local temperature exists in the shaded area.}
\label{temp_Si}
\end{figure}

Since $n\ind{min}$ has been calculated for a one dimensional model the results
we obtain here should be valid for one dimensional or at least quasi one dimensional
structures of the respective materials. Let us consider two examples:

Silicon is used in many branches of technology. In its crystalline form, it
has a Debye temperature of $\Theta \approx 645 \,$K and
its lattice constant is $a_0 \approx 2.4 \,${\AA}.
Using these parameters,
figure \ref{temp_Si} shows the minimal length-scale on which temperature can exist
in a one-dimensional
silicon wire as a function of global temperature.
Here, the accuracy parameters $\alpha$ (see eq.(\ref{e_range})) and $\delta$
(see below eq.(\ref{groundenergy}))
are chosen to be $\alpha=10$ and $\delta=0.01$. Local temperature exists in the shaded area.

%
%
%

Recently, carbon has been investigated for the fabrication of nano-structured devices
\cite{Dresselhaus1996,Dresselhaus2001}. In particular, we consider carbon nanotubes here,
which are widely used in nano-technological experiments.
Carbon nanotubes have diameters of only a few nanometers. Measurements of their specific heat
have shown, that their thermal properties can be accurately modeled with one-dimensional harmonic chains
\cite{Hone2002}. The presented results can thus be expected to be accurately applicable to them.
Carbon nanotubes have a Debye temperature of $\Theta \approx 1100 \,$K and a lattice constant
of $a_0 \approx 1.4 \,${\AA}.

%
%
%
\begin{figure}[h]
\psfrag{-1.1}{\small \raisebox{-0.1cm}{$10^{-1}$}}
\psfrag{1.1}{\small \raisebox{-0.1cm}{$10^{1}$}}
\psfrag{2.1}{\small \raisebox{-0.1cm}{$10^{2}$}}
\psfrag{3.1}{\small \raisebox{-0.1cm}{$10^{3}$}}
\psfrag{4.1}{\small \raisebox{-0.1cm}{$10^{4}$}}
\psfrag{5.1}{\small \raisebox{-0.1cm}{$10^{5}$}}
\psfrag{2}{\small \hspace{-0.7cm} $10^{-8}$}
\psfrag{4}{\small \hspace{-0.7cm} $10^{-6}$}
\psfrag{6}{\small \hspace{-0.7cm} $10^{-4}$}
\psfrag{8}{\small \hspace{-0.7cm} $10^{-2}$}
\psfrag{10}{\small \hspace{-0.0cm} $1$}
\psfrag{n}{\raisebox{0.1cm}{\hspace{-0.2cm}$l\ind{min}$ [m]}}
\psfrag{c1}{$\: T$ [K]}
\hspace{-0.16cm}
\includegraphics[width=7cm]{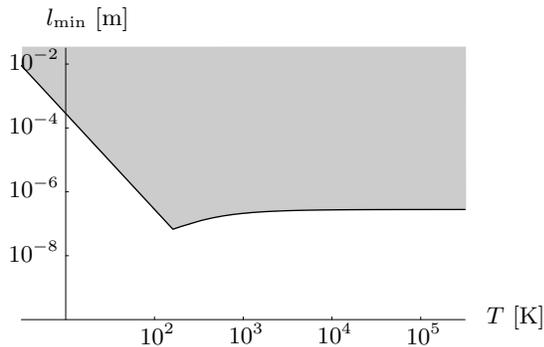}
\caption[$l\ind{min} (T)$ for a carbon nanotube.]{$l\ind{min}$ as a function of temperature $T$ for a
carbon nanotube. $a_0 \approx 1.4 \,${\AA},
$\Theta \approx 1100 \,$K, accuracy parameters $\alpha=10$ and $\delta=0.01$.
Local temperature exists in the shaded area.}
\label{temp_tube}
\end{figure}

Figure \ref{temp_tube} shows the minimal length-scale on which temperature can exist in a carbon nanotube
as a function of global temperature. It provides a good estimate of the maximal accuracy, with which
temperature profiles in such tubes can be meaningfully discussed \cite{Cahill2003}.
Again, the accuracy parameters $\alpha$ and $\delta$
\index{accuracy parameter}
are chosen to be $\alpha=10$ and $\delta=0.01$. Local temperature exists in the shaded area.

Of course the validity of the harmonic lattice model will eventually break down at high but finite
temperatures. The estimates drawn from the considered approach,
in particular the results presented in
figures \ref{temp_Si} and \ref{temp_tube}, will then no longer apply.


%
\section{Discussion of the Length Scale Results}
\label{discussion}

The length scales one obtains here are, in particular for low temperatures, surprisingly
large. One might thus wonder whether the approach really captures the relevant physics.
Let us therefore discuss some possible limitations: 

Firstly, one may argue that taking the limit of an infinite number of groups, as required for the
central limit theorem, will not correspond to the physically relevant situations.
However, having in mind that we intended to analyze when a small part of a larger system
can be in a thermal state, taking this limit should be well justified.

Secondly, only one-dimensional models were considered. A real physical system, even if
it is of a very prolate shape, is always three-dimensional. A generalization of the
approach to those models is thus of high interest. However let us stress here, that the general
conditions (\ref{cond_const}) and (\ref{cond_linear_1}) apply to systems of arbitrary
dimension, it is only the application to specific models, which needs to be generalized.

Furthermore, the harmonic chain is an exactly diagonalizable model,
this means that no phonon scattering does occur.
The purely harmonic model does, for example,
not predict any expansion or shrinking of the material caused
by heating or cooling. It is therefore possible that the harmonic
model may fail to give reliable results for our present investigation.
In particular at low temperatures, entanglement \cite{PlenioVedral1998,Nie98,AEP+02}
plays an important role and this effect
can be highly non-linear. 

Finally, one might speculate whether the length scales could significantly change if the
assumption of a global equilibrium state was relaxed. This possibility of course exists,
nonetheless one would expect the estimates to still apply as long as temperature
gradients are small. Imagine, there are two baths attached to the ends of the considered
harmonic chain of
section \ref{harmonicchain}. If both baths have the same temperature, the chain is
in a ``global'' equilibrium state and the present results are applicable.
If one now continously increased the temperature
of one bath, the density matrix of the chain would change continously, too.
Hence, the minimal group sizes
would also change continuously and the present results should still be good estimates,
at least for small temperature gradients.

To clarify whether the above findings are in agreement or in conflict with experiments,
their measurability needs to be considered in more detail. We proceed to do this in the
next section.


%
%
\section{Consequences for Measurements}
\label{measure}

In this section, we give some examples of possible experimental consequences of the local
breakdown of the temperature concept at small length scales, i.e. of the fact that the
respective individual subsystems or even subgroups of those do not reach
a canonical state.

\subsection{Standard Temperature Measurements}

Temperature is always measured indirectly via observables, which, in quantum mechanics,
are represented by hermitian operators. Usually, one is interested in measuring the
temperature of a system in a stationary state. The chosen observable should therefore be
a conserved quantity, i.e. its operator should commute with the Hamiltonian of the
system.

A conventional technique, e.g., is to bring the piece of matter, the temperature
$T$ of which is to be measured, in thermal contact with a box of an ideal
gas and to measure the pressure $p$ of the gas, which is related to its temperature
by $n \, k\ind{B} \, T = p \, V$ (cf. sect. \ref{sec:intro}).
Since the gas is in thermal equilibrium with the considered piece of matter, both
substances have the same temperature. A measurement of $p$ for constant $V$ allows to
infer the global temperature $T$ of the piece of matter.

One might wonder, whether a small (possibly even nanoscopic) thermometer \cite{Gao2002},
which is locally coupled
to one subsystem of the large chain considered in section \ref{general}, is capable of measuring
a local temperature or whether the measurement would show any indications of a
possible local breakdown of temperature.

A prerequisite for the above gas thermometer to work properly is that the thermometer
does not significantly
perturb the system.
For our class of models this means that the thermometer system should only be weakly
coupled to the respective
subsystem of the chain and that it should be significantly smaller than the latter.
These two requirements ensure that the energy exchange between system and thermometer
would not significantly alter the energy contained in the system.
Therefore, this measurement scenario can be accurately modeled as follows:

Let the thermometer be represented by a single spin, which is locally coupled to a
harmonic chain, say. Since the coupling is assumed to be weak
and the chain is assumed to be very large
and in a thermal state, the present scenario can accurately be modeled with a
master equation approach \cite{Weiss1999}.
However, it is a well known result of such system bath models, that the
reduced density matrix of the spin relaxes into a canonical state with the temperature
being equal to the global temperature of the harmonic chain (i.e. the bath).
The spin (thermometer) thus measures the global temperature of the total chain,
even for perfectly local coupling.

As long as the chain is in a global equilibrium state, a temperature measurement of
this type thus does not have any spatial resolution at all.
It is only capable of measuring the global temperature of the chain.
Neither can any local temperatures be measured nor any signatures of their breakdown be detected.

This conclusion obviously cannot hold anymore for scenarios with only local but no
global equilibrium. Macroscopic temperature profiles are routinely measured, with the
standard technique described above.
Whether such measurements of temperature profiles are still possible for much smaller
systems and what their maximally possible spatial resolution is in that
case should be subject of further investigations.

According to the above considerations, one might think that the question of
local temperatures for systems in global equilibrium was an irrelevant
issue since it has no observable consequences. This, however is not the case.
In the following we turn to discuss an
example of such measurable consequences of the local breakdown of the concept of temperature.

\subsection{Non-thermal Local Properties}

We now turn to observables of the object (chain) itself, which could be used to infer local
temperatures
$T\ind{loc}$, i.e. temperatures of subsystems, provided the subsystems are in a canonical state.
On the other hand, if the respective subsystems are not in a canonical state, this fact
should then modify the measurement results for those observables.

The minimal group sizes calculated in section \ref{general} depend on the
global temperature and on the
strength of the interactions between neighboring subsystems.
Furthermore, local temperatures can even
exist for single subsystems if these are finite dimensional. In the limiting case of infinite
temperature, the density matrix of a chain of finite dimensional subsystems is proportional
to the identity matrix and thus has the same form in every basis including the product basis,
which in turn implies that local temperature would then exist for single subsystems
\cite{Hartmann2004a}.
For systems composed of finite dimensional subsystems, local temperatures do thus exist
for single subsystems
at relatively low global temperatures if the coupling is weak, while they do not if the
coupling is strong.

Pertinent systems, for which such effects could easily by studied, are magnetic materials
\cite{Hartmann2004c}.
These can in many cases be described by spin lattice or spin chain models.
Since, as we will see below, properties of single spins can be infered from measurements
of even macroscopic magnetic observables, those materials thus allow
to study the existence of temperature,
as defined by the existence of a canonical state, on the most local scale possible,
i.e. for single spins.

For a spin-1/2 system, it is always possible to assign a Boltzmann factor and thus
a local temperature to the ratio of the occupation probability of the higher and lower
level. Here we consider a homogeneous chain of spin-1 particles interacting
with their nearest neighbors. For the interactions, one assumes a Heisenberg model.
The Hamiltonian of this
system reads \cite{vanVleck1945}:
\begin{equation} \label{measure_eq:1}
H = B \, \sum_{j=1}^n \sigma_j^z +
J \, \sum_{j=1}^n \sigma_j^x \sigma_{j+1}^x + \sigma_j^y \sigma_{j+1}^y +
\sigma_j^z \sigma_{j+1}^z \, ,
\end{equation}
where $\sigma_j^x$, $\sigma_j^y$ and $\sigma_j^z$ are the
spin-$1$ matrices. 
$B$ is an applied magnetic field, $J$ the coupling and $n$ the number of spins.
The coupling $J$ is taken to be positive, $J > 0$. The spins thus tend to align
anti-parallelly and the material is anti-ferromagnetic.\index{anti-ferromagnet}
The local Hamiltonian of subsystem $j$ is $H_j = B \, \sigma_j^z$.
The system has periodic boundary conditions and is thus translation invariant.
As in the previous sections, the entire system (\ref{measure_eq:1})
is assumed to be in a thermal state (see equation (\ref{can_def})).

As an example of an experiment, we will now consider two different magnetic
observables of a spin-$1$ system with the Hamiltonian (\ref{measure_eq:1}).
The first observable is the magnetization in the direction of the applied field,
$m_z$, which is here defined to be the total magnetic moment per particle: 
\begin{equation} \label{measure_eq:2}
m_z \equiv \frac{1}{n} \Big\langle \sum_{j=1}^n \sigma_j^z \Big\rangle \, ,
\end{equation}
where $\langle \mathcal{O} \rangle$ is the expectation value of the operator $\mathcal{O}$,
i.e. $\langle \mathcal{O} \rangle = \textrm{Tr}(\rho \, \mathcal{O})$.
In the translation invariant state $\rho$, the reduced
density matrices of all individual spins are equal, and the magnetization (\ref{measure_eq:2})\index{magnetization}
can be written as
\begin{equation} \label{measure_eq:3}
m_z = \langle \sigma_k^z \rangle \, ,
\end{equation}
for any $k = 1, 2, \dots, n$.
The magnetization, although defined macroscopically, is thus actually a
property of a single spin, i.e. a strictly local property.

As a second observable one can choose the occupation probability, $p$, of the $s_z = 0$
level (averaged over all spins),
\begin{equation} \label{measure_eq:5}
p = \frac{1}{n} \Big\langle \sum_{j=1}^n \left\ve 0_j \rangle \langle 0_j \right\ve \Big\rangle \, .
\end{equation}
Similar to $m_z$ according to equation (\ref{measure_eq:3}), $p$ may be written as
\begin{equation} \label{measure_eq:4}
p = \langle \left\ve 0_k \rangle \langle 0_k \right\ve \rangle \, ,
\end{equation}
for any $k = 1, 2, \dots, n$ and is thus strictly local, too.

Now, if each single spin was in a canonical state with temperature
$T\ind{loc}$,
$m_z$ and $p$ would both have to be monotonic functions of $T\ind{loc}$.
In this case, $T\ind{loc}$ could, after calibration,
be infered from measurements of $m_z$ or $p$, respectively. Note, that
$m_z$ is proportional to the local energy, the average energy of one subsystem.

\begin{figure}
\centering
\psfrag{-0.2}{\small \raisebox{-0.0cm}{\hspace{-0.2cm}$-0.2$}}
\psfrag{-0.4}{\small \raisebox{-0.0cm}{\hspace{-0.2cm}$-0.4$}}
\psfrag{-0.6}{\small \raisebox{-0.0cm}{\hspace{-0.2cm}$-0.6$}}
\psfrag{-0.8}{\small \raisebox{-0.0cm}{\hspace{-0.2cm}$-0.8$}}
\psfrag{-1.1}{\small \raisebox{-0.0cm}{\hspace{-0.2cm}$-1.0$}}
\psfrag{0}{\small \raisebox{-0.1cm}{$0$}}
\psfrag{2}{\small \raisebox{-0.1cm}{$2$}}
\psfrag{4}{\small \raisebox{-0.1cm}{$4$}}
\psfrag{6}{\small \raisebox{-0.1cm}{$6$}}
\psfrag{8}{\small \raisebox{-0.1cm}{$8$}}
\psfrag{0.4}{\small \raisebox{-0cm}{\hspace{0.1cm}$0.4$}}
\psfrag{0.3}{\small \raisebox{-0cm}{\hspace{0.1cm}$0.3$}}
\psfrag{0.2}{\small \raisebox{-0cm}{\hspace{0.1cm}$0.2$}}
\psfrag{0.1}{\small \raisebox{-0cm}{\hspace{0.1cm}$0.1$}}
\psfrag{1.2}{\small \raisebox{-0cm}{\hspace{0.1cm}$0.0$}}
\psfrag{m}{\hspace{-0.05cm}\raisebox{0.3cm}{$m_z$}}
\psfrag{p}{\hspace{0.05cm}\raisebox{-0.3cm}{$p$}}
\psfrag{T}{\hspace{-0.24cm}\raisebox{-0.3cm}{$\: T / B$}}
\includegraphics[width=7cm]{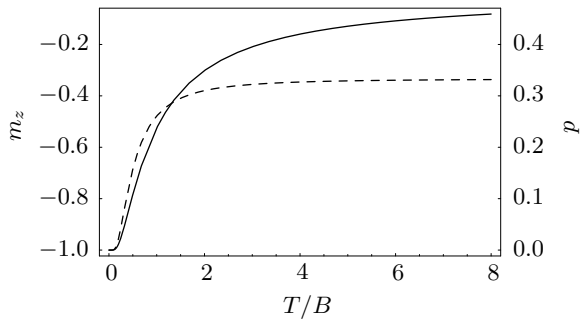}
\caption[$m_z(T)$ and $p(T)$ for the spin-1 chain with $J = 0.1 \times B$.]{\label{measure_fig:1} $m_z$ (solid line) and $p$ (dashed line) as a function of
temperature $T$ for a spin-$1$ chain of 4 particles.
$T$ is given in units of $B$ and $J = 0.1 \times B$.}
\end{figure}

Figure \ref{measure_fig:1}, shows $m_z$ and $p$ as a function
of the global temperature $T$ for
a spin-$1$ chain of 4 particles with the Hamiltonian (\ref{measure_eq:1})
for weak interactions, $J = 0.1 \times B$.
Both quantities are monotonic functions of each other.
%

The situation changes drastically when the spins are strongly coupled.
In this case the concept of temperature breaks down locally due to
correlations of each single spin with its environment.

\begin{figure}
\centering
\psfrag{-0.05}{\hspace{-0.02cm} \small \raisebox{-0.0cm}{$ 0.00$}}
\psfrag{-0.02}{\small \raisebox{-0.0cm}{$ $}}
\psfrag{-0.04}{\hspace{-0.3cm} \small \raisebox{-0.0cm}{$-0.04$}}
\psfrag{-0.06}{\small \raisebox{-0.0cm}{$ $}}
\psfrag{-0.08}{\hspace{-0.3cm} \small \raisebox{-0.0cm}{$-0.08$}}
\psfrag{-0.1}{\small \raisebox{-0.0cm}{$ $}}
\psfrag{-0.12}{\hspace{-0.3cm} \small \raisebox{-0.0cm}{$-0.12$}}
\psfrag{-0.14}{\small \raisebox{-0.0cm}{$ $}}
\psfrag{0}{\small \raisebox{-0.1cm}{$0$}}
\psfrag{2}{\small \raisebox{-0.1cm}{$2$}}
\psfrag{4}{\small \raisebox{-0.1cm}{$4$}}
\psfrag{6}{\small \raisebox{-0.1cm}{$6$}}
\psfrag{8}{\small \raisebox{-0.1cm}{$8$}}
\psfrag{0.332}{\small \raisebox{-0cm}{\hspace{0.1cm}$0.332$}}
\psfrag{0.334}{\small \raisebox{-0cm}{$ $}}
\psfrag{0.336}{\small \raisebox{-0cm}{\hspace{0.1cm}$0.336$}}
\psfrag{0.338}{\small \raisebox{-0cm}{$ $}}
\psfrag{0.34}{\small \raisebox{-0cm}{\hspace{0.1cm}$0.340$}}
\psfrag{0.342}{\small \raisebox{-0cm}{$ $}}
\psfrag{0.344}{\small \raisebox{-0cm}{\hspace{0.1cm}$0.344$}}
\psfrag{0.346}{\small \raisebox{-0cm}{$ $}}
\psfrag{m}{\hspace{0.0cm}\raisebox{0.3cm}{$m_z$}}
\psfrag{p}{\hspace{-0.1cm}\raisebox{-0.3cm}{$p$}}
\psfrag{T}{\hspace{-0.24cm}\raisebox{-0.24cm}{$\: T / B$}}
\includegraphics[width=7cm]{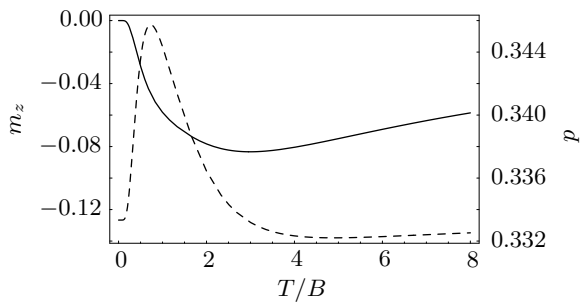}
\caption[$m_z(T)$ and $p(T)$ for the spin-1 chain with $J = 2 \times B$.]{\label{measure_fig:3} $m_z$ (solid line) and $p$ (dashed line) as a function of
temperature $T$ for a spin-$1$ chain of 4 particles.
$T$ is given in units of $B$ and $J = 2 \times B$.}
\end{figure}

Figure \ref{measure_fig:3} shows $m_z$ and $p$ as a function of temperature $T$ for
a spin-$1$ chain of 4 particles with the Hamiltonian (\ref{measure_eq:1})
for strong interactions $J = 2 \times B$.
Both quantities are non-monotonic functions of $T$ and therefore no mapping
between $m_z$ and $p$ exists.
%

How could a local observer determine whether the system he observes, a single spin,
is in a thermal state and can therefore be characterized by a temperature?
The local observer would need to compare two situations:

In the first situation, the spin is weakly coupled to a larger system, the heat bath.
In this situation, the local observer could measure $m_z$ and $p$ as functions of 
the temperature of the heat bath and would get a result similar to figure \ref{measure_fig:1}.
This result would not be sensitive to the details of the coupling to the
heat bath. The local observer would thus recognize this situation as a particular one
and might term it the ``thermal'' situation.

The second situation is fundamentally different. The spin is now strongly coupled to
its surrounding. If the local observer again measures $m_z$ and $p$ as functions of 
the temperature of the surrounding, he would get a result like in figure \ref{measure_fig:3}.

The observer can tell the difference between both situations, even if he has no access to the
global (true) temperature $T$ of the surrounding.
In the first case he can construct a mapping from
say $m_z$ to $p$, i.e. $p(m_z)$, or vice versa, $m_z(p)$, in the second he cannot:
There exist for example two values of $p$ corresponding to only one value of $m_z$.
Here the concept of a local temperature breaks down
at least on the level of individual particles, since temperature measurements
via different local observables would contradict each other.

The question, whether and on what scale local temperatures can exist in systems,
which are in a global
equilibrium state is thus indeed physically relevant. The advantage of the concept of temperature
is that it allows to predict various physical properties of the considered system.
This is only possible if different properties (expectation values of observables)
map one to one on
each other as in figure \ref{measure_fig:1}. The following example illustrates the situation:

Consider a piece of metal, say a wire. Assume , its temperature is measured via its
electrical resistance.
Why are we interested in this temperature? We are interested in it because it also
allows us to predict how the wire behaves with respect to other physical processes.
For example, if we know its temperature,
we can tell whether the wire is going to melt or not. Effectively, we thus have a mapping
of the resistance
onto the fact that the wire is going to melt or is not going to melt.
In a more mathematical language, we can construct a function: melting as a function of the
resistance.
Analogously, for the scenario of figure \ref{measure_fig:1}, a local observer
is able to construct a function
$m_z (p)$.

What happens if such functions can no longer be constructed? In this situation,
the concept of temperature
becomes useless. Assume our wire had such properties. We could still measure its
resistance and, if we
wished, could assign a ``temperature value'' to it. This ``temperature value'',
however, would be of no further use,
since it would not allow us to predict whether the wire is going to melt or not.  
A situation where such problems do really occur is the scenario of figure \ref{measure_fig:3}.

\subsection{Potential Experimental Tests}

Finally, we address the question of whether the effects described here could be observed
in real experiments. Indeed, pertinent experiments are available and have partly
already been carried out:

A realization of a quasi one dimensional anti-ferromagnetic spin-1 Heisenberg chain is
the compound CsNiCl$_3$\index{CsNiCl$_3$}
\cite{Kenzelmann2001,Affleck1989,Avenel1992}. Here the coupling is\linebreak
$J \approx 2.3$~meV. To achieve a detectable modulation of $m_z$ and $p$, the spins
should be significantly polarized for $T > 0$.
Therefore a sufficiently strong applied magnetic field is needed.
For CsNiCl$_3$, a field of roughly $9.8$~Tesla would correspond to $J = 4 \times B$.

The magnetization in an applied field can be measured with high precision by means of a
SQUID \cite{Lipa1981}. The occupation probability of the $s_z = 0$ states, on the other
hand, is accessible via neutron scattering experiments
\cite{Kenzelmann2002a,Ma1995}.
The differential cross section for neutron scattering of spin-systems is 
a function of the Fourier transforms of spin correlation functions \cite{Lovesey1989}.
One can thus obtain information about the quantity
\begin{equation}
\frac{1}{n} \, \sum_{\vec{r}}
\langle \sigma^x_{\vec{r}}(0) \, \sigma^x_{\vec{r}}(0) +
\sigma^y_{\vec{r}}(0) \, \sigma^y_{\vec{r}}(0) \rangle = 1 + p
\end{equation}
from the measurement data.
Therefore, $p$ is measurable in neutron scattering experiments.

Such experiments or a combination thereof could thus be used to demonstrate the
non-existence of local temperature.



%
\section{Conclusion and Outlook} \label{conclusion}

In the present work, we have considered the minimal spatial length
scales on which local temperature can meaningfully be
defined. For large systems in a global equilibrium state,
we have reviewed the derivation of two criteria which are valid for
quantum many body systems with nearest
neighbor interactions and discussed the physical relevance of the existence
and non-existence of local temperatures. 

Some questions related to the microscopic limit of the applicability of thermodynamics
have thus been clarified. Nevertheless, some open problems remain and
even new ones appeared in the context of the present approach.

First of all, the generalization of the calculations to scenarios with only local but no global
equilibrium is an issue of significant importance.
One might expect that the length scales do no longer
depend on the interactions and the global temperature only,
but that the temperature gradient becomes
relevant, too. 

For global non-equilibrium, local temperature measurements of the standard type are
very interesting
and important issues on their own. As we have discussed here, these measurements
have no spatial resolution
if the sample is in a global equilibrium state. On the other hand,
local temperature measurements with
spatial resolution are being done for macroscopic setups and nobody would
dare to question their validity.
Therfore the maximal spatial resolution of this kind of measurements is an
interesting question and the
present understanding of this topic is quite poor.

Future research could also be concerned with new physics
that might show up for small
entities, which are in contact with a thermal surrounding,
but show non-thermal behavior due to
the breakdown of temperature on the respective scale. One example for this are the observable
features discussed in section \ref{measure}. However, one might think about more surprising
phenomena, as for example anomalous pressure fluctuations in very small gas bubbles enclosed
in a piece of solid. With respect to future nanotechnologies,
such phenomena could equally be harmful
or useful, depending on whether one is able to design the devices in the pertinent way. 

Finally, possible generalizations of thermodynamics, that could apply on even smaller scales
seem to be interesting. In the present work, we have considered the microscopic limit of usual
thermal behavior in quantum systems, i.e. Quantum Thermodynamics \cite{Gemmerbuch},
where effective interactions among the considered parts are small.
One might thus wonder, whether only partitions with weak effective couplings can be considered
within such a ``universal'' description, that does not depend on the details of the
microscopic constituents,
or whether there exists again an intermediate level of description, not as universal as standard
thermodynamics but applicable on smaller scales.
Since, in standard thermodynamics, equilibrium states are fully characterized by
one single parameter, temperature (cf. equation (\ref{can_def})),
one could for example imagine that there
exists a class of generalized equilibrium states,
which require say two or three parameters for their
characterization. Some phenomenological attempts in this direction have already been made
\cite{Hill2001c,Rajagopal2004}.
Nonetheless, a justification of these attempts from an underlying theory, i.e. quantum or
classical mechanics,
is still missing.


\begin{acknowledgements}
The author likes to thank G\"unter Mahler very much for proof-reading the manuscript and for
many helpfull comments.
\end{acknowledgements}


\bibliographystyle{unsrt}
\bibliography{../BIB/mybib}



\end{document}